# Velocity field within a vortex ring with a large elliptical cross section

BY T. S. MORTON

The velocity field within a steady toroidal vortex is found for arbitrary mean core radius and section ellipticity. The problem is solved by transforming to coordinates that define invariant sets. The method allows the properties of the coordinate system metric tensor to be exploited in the continuity equation in order to obtain the solution. The vorticity is found to decrease monotonically with distance from the symmetry axis. For a given outer radius and outer perimeter velocity, the circulation of the vortex ring can be either smaller or larger than that of Hill's spherical vortex.

## 1. Introduction

The toroidal vortex can be useful in studying a variety of problems in science and engineering, such as axisymmetric wakes, jets, and electromagnetic phenomena. Analytical solutions for vortex rings have invariably involved Bessel functions and asymptotic expansions (see e.g., Fukumoto and Moffatt 2000) in order to describe the velocity distribution. These types of solutions are restricted to rings of small cross section, which limit the types of studies that can be made of the vortex core itself. In the present work, an explicit algebraic expression is found for the velocity field within the core of a vortex ring with a large elliptical cross section.

One of the first solutions ever found for axisymmetric vortex regions was the spherical vortex of Hill (1894). There are several accounts of this solution (Lamb 1932; Milne-Thomson 1968; Panton 1984; Saffman 1992). Extensions of Hill's spherical vortex were found by O'Brien (1961), by allowing the sphere to distort into an ellipsoid. A distinctive feature of Hill's spherical vortex is the bounded velocity on the axis of symmetry. Another axisymmetric vortex structure is the vortex ring (see Kelvin 1869, Maxwell 1873, Lamb 1932), in which the entire field is irrotational except for a ring-shaped rotational core whose diameter is small relative to its distance from the axis of symmetry. This small, ring-shaped core is assumed to have constant vorticity. Lichtenstein (1925) studied a vortex ring of finite cross section and uniform vorticity. He found that for small cross section, the shape of the section approximates an ellipse, with the minor axis parallel to the axis of symmetry. The existence of steady vortex rings in an inviscid fluid has been proved by Fraenkel (1970) and Maruhn (1957). Examples of such vortex rings of small cross section are given by Fraenkel (1972) using expansions in powers and logarithms of a small cross section parameter. Saffman (1970) derived a formula for the propagation velocity of a vortex ring (circular to leading order) in an ideal fluid with an arbitrary distribution of vorticity in the core. Norbury (1972) determined a function describing the core boundary of a family of vortex rings for a prescribed vorticity distribution and velocity of propagation by writing the formal solution as an integral equation and solving the problem numerically. Norbury (1973) introduced a parameter, $\alpha$, termed the "non-dimensional mean core radius" which varies in the range $0 < \alpha \leq \sqrt{2}$. It can represent a family of vortices that include vortex rings of small cross section, where $\alpha \to 0$, as well as Hill's spherical vortex, where $\alpha = \sqrt{2}$. Hill's spherical vortex was considered to be a limiting case of a family of vortex rings that are possible with a vorticity distribution given by $\omega(3)/R$, where $\omega(3)$ is equal to the physical component of vorticity in the direction circling the symmetry axis. This ratio, which was taken as constant, has been referred to as a "vorticity constant" (Norbury 1973) as well as a "vorticity density" (Mohseni & Gharib 1998; Mohseni 2001). Hunt and Eames (2002) stated that during axisymmetric stretching in a straining flow the quantity $\omega(3)/R$ is conserved and that it is proportional to the circulation around a vortex element. What is seldom, if ever, mentioned, however, is that this constant is actually the contravariant component of the vorticity tensor, $\tilde{\omega}^k$, referenced to a spherical coordinate system defined by

$$x \equiv x^1 = \tilde{x}^1 \sin(\tilde{x}^2)\cos(\tilde{x}^3)$$

$$y \equiv x^2 = \tilde{x}^1 \sin(\tilde{x}^2)\sin(\tilde{x}^3)$$

$$z \equiv x^3 = \tilde{x}^1 \cos(\tilde{x}^2).$$

Here, $\tilde{x}^1 = \sqrt{x^2 + y^2 + z^2}$, $\tilde{x}^2 = \theta$, and $\tilde{x}^3 = \varphi$. A diagram of the coordinate system is shown in Figure 1. The metric tensor, $\tilde{g}_{ij}$, relating the spherical coordinate system to a rectangular system is given by:

$$\tilde{g}_{ij} = \begin{bmatrix} 1 & 0 & 0 \\ 0 & (\tilde{x}^1)^2 & 0 \\ 0 & 0 & (\tilde{x}^1)^2 \sin^2(\tilde{x}^2) \end{bmatrix}.$$

To see that the vorticity tensor is uniform, we express the vorticity definition in the spherical coordinate system as:

$$\tilde{\omega}^i = \tilde{E}^{ijk} \tilde{v}_{k,j} = \frac{\varepsilon^{ijk}}{\sqrt{\tilde{g}}} \tilde{v}_{k,j}.$$

Here, $\varepsilon^{ijk}$ is the permutation symbol and $\tilde{E}^{ijk}$ the permutation tensor in the spherical coordinate system. Setting $i = 3$ and noting the orthogonality of the coordinate system, the above definition becomes:

$$\tilde{\omega}^3 = \frac{1}{\sqrt{\tilde{g}}} \left[ \frac{\partial}{\partial \tilde{x}^1} \tilde{g}_{22}\tilde{v}^2 - \frac{\partial}{\partial \tilde{x}^2} \tilde{g}_{11}\tilde{v}^1 \right].$$

Employing the Stokes stream function and the fact that the physical component, $\tilde{v}(2)$, of velocity in the $\tilde{x}^2$ direction is $\tilde{v}(2) = \sqrt{\tilde{v}_2 \tilde{v}^2} = \sqrt{\tilde{g}_{22}} \tilde{v}^2 = \tilde{x}^1 \tilde{v}^2$ then gives:

$$\tilde{\omega}^3 = -\frac{1}{(\tilde{x}^1)^2 \sin(\tilde{x}^2)} \left[ \frac{\partial}{\partial \tilde{x}^1} \left( \frac{1}{\sin(\tilde{x}^2)} \frac{\partial \psi}{\partial \tilde{x}^1} \right) + \frac{\partial}{\partial \tilde{x}^2} \left( \frac{1}{(\tilde{x}^1)^2 \sin(\tilde{x}^2)} \frac{\partial \psi}{\partial \tilde{x}^2} \right) \right].$$

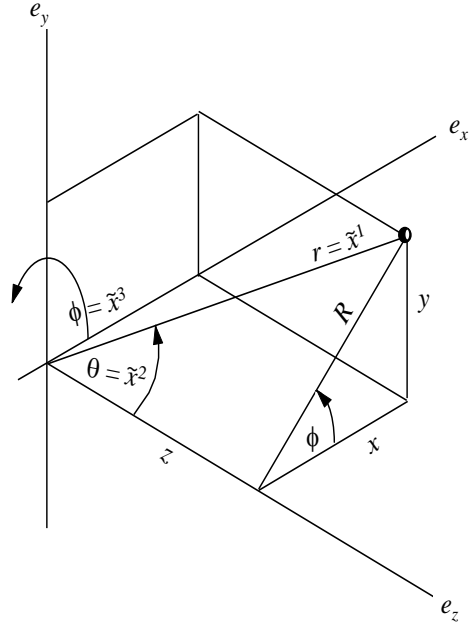

Figure 1. Spherical coordinate system used for Hill's spherical vortex.

Assuming a uniform vorticity distribution in the $\tilde{x}^3$ direction and a product solution of the form: $\psi = F(\tilde{x}^1)\sin^2(\tilde{x}^2)$ yields the Cauchy-Euler equation $(\tilde{x}^1)^2 F'' - 2F = -(\tilde{x}^1)^4 \tilde{\omega}^3$. This is then solved, and the requirement that velocities remain bounded as $\tilde{x}^1 \to 0$ is imposed, giving:

$$\psi = \tilde{\omega}^3 \frac{\left[R_O^{\,2}(\tilde{x}^1)^2 - (\tilde{x}^1)^4\right]}{10} \sin^2(\tilde{x}^2).$$

where $\psi = 0$ when $\tilde{x}^1 = R_O$. The physical components of the velocity are then found to be:

$$\tilde{v}(1) = \frac{1}{(\tilde{x}^1)^2 \sin(\tilde{x}^2)} \frac{\partial \psi}{\partial \tilde{x}^2} = \tilde{\omega}^3 \frac{\left[R_O^{\,2} - (\tilde{x}^1)^2\right]}{5} \cos(\tilde{x}^2) \tag{1a}$$

$$\tilde{v}(2) = \frac{-1}{\tilde{x}^1 \sin(\tilde{x}^2)} \frac{\partial \psi}{\partial \tilde{x}^1} = \tilde{\omega}^3 \frac{\left[2(\tilde{x}^1)^2 - R_O^{\,2}\right]}{5} \sin(\tilde{x}^2). \tag{1b}$$

Setting $\tilde{v}(1) = 0$, $\tilde{v}(2) = 0$ locates the vortex centre at $\tilde{x}^1 = R_O/\sqrt{2}$, $\tilde{x}^2 = \pi/2$.

To summarize, in terms of the tensor components of the spherical coordinate system, Hill's spherical vortex is a finite three-dimensional region of uniform vorticity, the vorticity components being $\tilde{\omega}^1 = 0$, $\tilde{\omega}^2 = 0$, and

$$\tilde{\omega}^3 = 5v_O / R_O^{\,2}, \tag{2}$$

and the quantity $v_O$ being defined as $\tilde{v}(2)$ at $\tilde{x}^1 = R_O$, $\tilde{x}^2 = \pi/2$. The physical component, $\tilde{\omega}(3)$, of $\tilde{\omega}^3$ is given by:

$$\tilde{\omega}(3)_{\text{Hill}} = \sqrt{\tilde{\omega}_3\,\tilde{\omega}^3} = \sqrt{\tilde{g}_{33}}\tilde{\omega}^3 = \tilde{x}^1 \sin(\tilde{x}^2)\,\tilde{\omega}^3 = R\,\tilde{\omega}^3. \tag{3}$$

which is the well known linear dependence.

Therefore, it is only the physical component of vorticity that varies linearly with distance from the symmetry axis. Hill's spherical vortex is a sphere of uniform vorticity and is analogous to the circular or elliptical patch of uniform vorticity in planar flow.

Yarmitskii (1975) outlined an extension of Hill's solution in which an azimuthal velocity component $\tilde{v}(3)$ arises inside the spherical vortex. Berezovskii and Kaplanskii (1987) obtained a toroidal vorticity distribution in the form of an asymptotic expansion. In their study, the modified Bessel function and the exponential functions appearing in their expression for the velocity were approximated by polynomials. For an excellent review of vortex rings, see Shariff and Leonard (1992).

For Hill's spherical vortex, the velocity at the outer radius of the vortex is equal in magnitude to the maximum return velocity along the axis of symmetry. For vortex rings, however, velocities can become much greater at the inner radius than $v_O$ at the outer radius. While the general shape and propagation speed of vortex rings are known from the studies mentioned above, detailed geometrical parameters such as $\alpha$ and $R_O/R_C$ can vary somewhat depending upon the flow configuration (see e.g. the experimental works of Gharib *et al.* (1998) and Southerland *et al.* (1991), Didden (1979), Magarvey and MacLatchy (1964), Van Dyke (1982), and Ober *et al.* (1995))). To date there has been no explicit algebraic expression for the velocity field within the core of a vortex ring of large elliptical cross section. The purpose of the present work is to find such an expression and to give a fairly general formula for the stream function.

## 2. Toroidal Vortex Solution

The general approach will be to introduce a toroidal coordinate system that places all motion into a single variable from a Lagrangian viewpoint. The coordinate system will define invariant sets in which the fluid particles must reside. This simplification, along with properties of the coordinate system metric tensor, allows the continuity equation to be manipulated into a relatively simple expression for the velocity distribution. This is then integrated round the perimeter of the vortex in order to extract an expression for the velocity field from the circulation.

### *2.1. Coordinate System*

The initial step in the solution is to construct a coordinate system in which contours of two of the coordinates coincide with streamlines of the flow. In this way, time derivatives of these will vanish, and the time derivative of the remaining coordinate completely describes the motion of the fluid. For the toroidal vortices that form in the wake of axisymmetric bodies, the following coordinate system (with coordinates $\bar{x}^i$), defined in terms of the components $\hat{x}^j$ of a cylindrical coordinate system, has the desired properties:

$$\tag{4}$$

$$\hat{x}^1 = z = \bar{x}^1 A \cos(\bar{x}^2)$$
$$\hat{x}^2 = R = \bar{x}^1 B \sin(\bar{x}^2) - k + R_C$$
$$\hat{x}^3 = \varphi = \bar{x}^3$$

The domain of interest is $(0 \leq \bar{x}^1 \leq \bar{x}^1_{\max})$ and $(0 \leq \bar{x}^2, \bar{x}^3 \leq 2\pi)$. This coordinate definition, shown in Figure 2, is similar to that used by Morton (1997) for the planar vortex, except for the presence of the constants $k$ and $R_C$. The constant $k$ allows the critical point in the vortex core to be shifted relative to the section centroid in a direction that is normal to the axis of symmetry. (The axis of symmetry lies parallel to the main flow.) The constant $R_C$ allows the entire vortex cross section to be displaced from the axis of symmetry. The reason for labelling the $\hat{x}^j$ coordinates as $z$, $R$, and $\varphi$ is because the present toroidal coordinate system is described relative to a cylindrical coordinate system (see Figure 2). Contours of $\bar{x}^1$ correspond to streamlines of the flow. The distance $\hat{x}^2 = R_C$ from the torus axis of symmetry to the critical point within the vortex cross section is found by setting $\bar{x}^1 = 0$. The constant $k$ can be evaluated by computing $\hat{x}^2$ on the outer streamline, that is, where $\bar{x}^1 = \bar{x}^1_{\max}$ and $\bar{x}^2 = -\pi/2$. (Note that the points on $\bar{x}^2 = \pm \pi/2$ lie on a line normal to the axis of symmetry.) This gives:

$$k = \frac{R_C - R_I - b}{b} \qquad \therefore (-1 < k < 1) \qquad (5)$$

where $b = \bar{x}^1_{\max} B$. (This definition preserves the area formula: $S = \pi ab$). Similar reasoning for the point $(\bar{x}^1_{\max}, \pi/2)$ yields the relation: $R_O = R_I + 2b$.

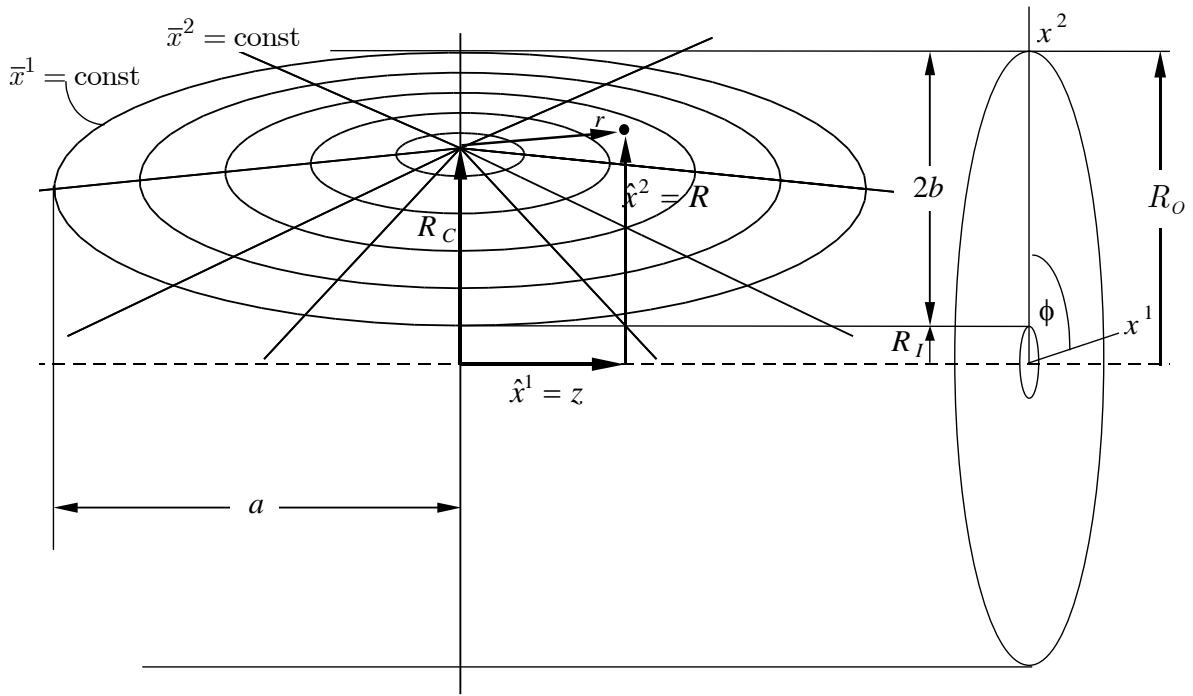

Figure 2. Diagram illustrating the construction of the toroidal coordinate system.

Defining the constants $A$ and $B$ as: $A = \cosh(q)$, $B = \sinh(q)$ for $q > 0$, allows the axis ratio of concentric ellipses to remain constant as $\bar{x}^1$ and $\bar{x}^2$ vary independently. It also ensures the useful identity:

$$A^2 - B^2 = 1. \tag{6}$$

Note that $q \to \infty$ corresponds to a circular vortex.

Considering now the coordinates $\hat{x}^j$ to be components of a cylindrical coordinate system, the transformation can be made to a Cartesian coordinate system, $x^k$, as follows:

$$\begin{aligned}
x^1 &= R \cos \phi = \hat{x}^2 \cos(\hat{x}^3) \\
x^2 &= R \sin \phi = \hat{x}^2 \sin(\hat{x}^3) \\
x^3 &= z = \hat{x}^1
\end{aligned} \tag{7}$$

This coordinate definition is also shown graphically in Figure 2.

The metric tensor $\bar{g}_{jq}$ relating the toroidal coordinate system to the rectangular system can be found by equating distance in the two coordinate systems, as follows:

$$ds^2 = \sum_{k=1}^{3} dx^k dx^k = \sum_{k=1}^{3} \frac{\partial x^k}{\partial \hat{x}^i} \frac{\partial \hat{x}^i}{\partial \bar{x}^j} d\bar{x}^j \frac{\partial x^k}{\partial \hat{x}^p} \frac{\partial \hat{x}^p}{\partial \bar{x}^q} d\bar{x}^q = \bar{g}_{jq} d\bar{x}^j d\bar{x}^q,$$

where

$$\bar{g}_{jq} \equiv \left( \sum_{k=1}^{3} \frac{\partial x^k}{\partial \hat{x}^i} \frac{\partial x^k}{\partial \hat{x}^p} \right) \frac{\partial \hat{x}^i}{\partial \bar{x}^j} \frac{\partial \hat{x}^p}{\partial \bar{x}^q} = \hat{g}_{ip} \frac{\partial \hat{x}^i}{\partial \bar{x}^j} \frac{\partial \hat{x}^p}{\partial \bar{x}^q} \tag{8}$$

and $\hat{g}_{ip}$ is the familiar metric tensor relating a cylindrical coordinate system to a rectangular system. In this case, its form is:

$$\hat{g}_{ip} = \begin{bmatrix} 1 & 0 & 0 \\ 0 & 1 & 0 \\ 0 & 0 & R^2 \end{bmatrix}.$$

Computing the partial derivatives in Eq. (8), the desired metric tensor $\bar{g}_{jq}$ becomes:

$$\bar{g}_{jq} = \begin{bmatrix} \left[A^2 \cos^2(\bar{x}^2) + B^2 \sin(\bar{x}^2) - k\right]^2 & \bar{x}^1 \cos(\bar{x}^2)\left[B^2 \sin(\bar{x}^2) - k - A^2 \sin(\bar{x}^2)\right] & 0 \\ & (\bar{x}^1)^2 \left[A^2 \sin^2(\bar{x}^2) + B^2 \cos^2(\bar{x}^2)\right] & 0 \\ & & R^2 \end{bmatrix}$$

Only the upper triangular elements of the symmetric metric tensor are shown. The Jacobian of the transformation from the toroidal coordinate system to the rectangular system can be found by computing the appropriate partial derivatives, as follows:

$$\frac{\partial x^i}{\partial \bar{x}^j} = \frac{\partial x^i}{\partial \hat{x}^k} \frac{\partial \hat{x}^k}{\partial \bar{x}^j} = \begin{bmatrix} \cos(\hat{x}^3) B [\sin(\bar{x}^2) - k] & \cos(\hat{x}^3) \bar{x}^1 B \cos(\bar{x}^2) & -\hat{x}^2 \sin(\hat{x}^3) \\ \sin(\hat{x}^3) B [\sin(\bar{x}^2) - k] & \sin(\hat{x}^3) \bar{x}^1 B \cos(\bar{x}^2) & \hat{x}^2 \cos(\hat{x}^3) \\ A \cos(\bar{x}^2) & -\bar{x}^1 A \sin(\bar{x}^2) & 0 \end{bmatrix}.$$

The Jacobian determinant $\sqrt{\bar{g}}$ is therefore:

$$\sqrt{\bar{g}} = \hat{x}^2 \, \bar{x}^1 AB[1 - k\sin(\bar{x}^2)], \tag{9}$$

and

$$\bar{g} = g_{33}(\bar{x}^1 AB)^2 [1 - k\sin(\bar{x}^2)]^2.$$

The inverse of the metric tensor is:

$$\bar{g}^{ij} = \frac{1}{\bar{g}} \begin{bmatrix} \bar{g}_{22}\bar{g}_{33} - \bar{g}_{23}\bar{g}_{32} & \bar{g}_{13}\bar{g}_{32} - \bar{g}_{12}\bar{g}_{33} & \bar{g}_{12}\bar{g}_{23} - \bar{g}_{13}\bar{g}_{22} \\ \bar{g}_{23}\bar{g}_{31} - \bar{g}_{21}\bar{g}_{33} & \bar{g}_{11}\bar{g}_{33} - \bar{g}_{13}\bar{g}_{31} & \bar{g}_{13}\bar{g}_{21} - \bar{g}_{11}\bar{g}_{23} \\ \bar{g}_{21}\bar{g}_{32} - \bar{g}_{22}\bar{g}_{31} & \bar{g}_{12}\bar{g}_{31} - \bar{g}_{11}\bar{g}_{32} & \bar{g}_{11}\bar{g}_{22} - \bar{g}_{12}\bar{g}_{21} \end{bmatrix}$$

However, due to the zero elements in $\bar{g}_{jq}$, the inverse of the metric tensor simplifies to:

$$\bar{g}^{ij} = \begin{bmatrix} \dfrac{\bar{g}_{22}\bar{g}_{33}}{\bar{g}} & \dfrac{-\bar{g}_{12}\bar{g}_{33}}{\bar{g}} & 0 \\ \dfrac{-\bar{g}_{21}\bar{g}_{33}}{\bar{g}} & \dfrac{\bar{g}_{11}\bar{g}_{33}}{\bar{g}} & 0 \\ 0 & 0 & \dfrac{1}{\bar{g}_{33}} \end{bmatrix} \tag{10}$$

## 2.2. Continuity Equation

Having constructed the streamline coordinate system, the velocity field can be found by applying the continuity equation in the new coordinate system:

$$\bar{v}^i{}_{,i} = 0. \tag{11}$$

Following the rules for covariant differentiation of the contravariant velocity vector in the toroidal coordinate system, the incompressible continuity equation is:

$$\frac{\partial \bar{v}^k}{\partial \bar{x}^k} + \bar{\Gamma}^k_{ik} \bar{v}^i = 0, \tag{12}$$

where $\bar{\Gamma}^k_{ij}$ are the Christoffel symbols of the second kind given by

$$\bar{\Gamma}^k_{ij} = \frac{\bar{g}^{kp}}{2}\left(\frac{\partial \bar{g}_{pi}}{\partial \bar{x}^j} + \frac{\partial \bar{g}_{pj}}{\partial \bar{x}^i} - \frac{\partial \bar{g}_{ij}}{\partial \bar{x}^p}\right). \tag{13}$$

The only nonzero velocity component in the toroidal coordinate system is $d\bar{x}^2/dt$ (recall that the curves $\bar{x}^1, \bar{x}^3 = \text{const.}$ are streamlines across which no fluid passes). With this simplification, the continuity equation (12) becomes:

$$\frac{\partial \bar{v}^2}{\partial \bar{x}^2} + \bar{\Gamma}^1_{21}\bar{v}^2 + \bar{\Gamma}^2_{22}\bar{v}^2 + \bar{\Gamma}^3_{23}\bar{v}^2 = 0. \tag{14}$$

Substituting (13) into (14) with $i = 2$, and $j$ and $k$ equal within each term, we have:

$$0 = \frac{\partial \bar{v}^2}{\partial \bar{x}^2} + \left[\frac{1}{2}\bar{g}^{11}\left(\frac{\partial \bar{g}_{12}}{\partial \bar{x}^1} + \frac{\partial \bar{g}_{11}}{\partial \bar{x}^2} - \frac{\partial \bar{g}_{21}}{\partial \bar{x}^1}\right) + \frac{1}{2}\bar{g}^{12}\left(\frac{\partial \bar{g}_{22}}{\partial \bar{x}^1} + \frac{\partial \bar{g}_{21}}{\partial \bar{x}^2} - \frac{\partial \bar{g}_{21}}{\partial \bar{x}^2}\right)\right]\bar{v}^2 \quad (j,k=1)$$

$$+ \left[\frac{1}{2}\bar{g}^{21}\left(\frac{\partial \bar{g}_{12}}{\partial \bar{x}^2} + \frac{\partial \bar{g}_{12}}{\partial \bar{x}^2} - \frac{\partial \bar{g}_{22}}{\partial \bar{x}^1}\right) + \frac{1}{2}\bar{g}^{22}\left(\frac{\partial \bar{g}_{22}}{\partial \bar{x}^2} + \frac{\partial \bar{g}_{22}}{\partial \bar{x}^2} - \frac{\partial \bar{g}_{22}}{\partial \bar{x}^2}\right)\right]\bar{v}^2 \quad (j,k=2)$$

$$+ \left[\frac{1}{2}\bar{g}^{33}\left(\frac{\partial \bar{g}_{32}}{\partial \bar{x}^3} + \frac{\partial \bar{g}_{33}}{\partial \bar{x}^2} - \frac{\partial \bar{g}_{23}}{\partial \bar{x}^3}\right)\right]\bar{v}^2 \quad (j,k=3)$$

By the symmetry in the two indices of both $\bar{g}^{ij}$ and $\bar{g}_{ij}$, the continuity equation simplifies to:

$$\frac{\partial \bar{v}^2}{\partial \bar{x}^2} + \left[\frac{1}{2}\bar{g}^{11}\left(\frac{\partial \bar{g}_{11}}{\partial \bar{x}^2}\right) + \bar{g}^{21}\left(\frac{\partial \bar{g}_{12}}{\partial \bar{x}^2}\right) + \frac{1}{2}\bar{g}^{22}\left(\frac{\partial \bar{g}_{22}}{\partial \bar{x}^2}\right) + \frac{1}{2}\bar{g}^{33}\left(\frac{\partial \bar{g}_{33}}{\partial \bar{x}^2}\right)\right]\bar{v}^2 = 0$$

The indices of $\bar{g}^{ij}$ are then lowered using Eq. (10) to give:

$$\frac{\partial \bar{v}^2}{\partial \bar{x}^2} + \frac{\bar{v}^2}{2\bar{g}}\left[\bar{g}_{22}\bar{g}_{33}\left(\frac{\partial \bar{g}_{11}}{\partial \bar{x}^2}\right) - 2\bar{g}_{12}\bar{g}_{33}\left(\frac{\partial \bar{g}_{12}}{\partial \bar{x}^2}\right) + \bar{g}_{11}\bar{g}_{33}\left(\frac{\partial \bar{g}_{22}}{\partial \bar{x}^2}\right) + \frac{\bar{g}}{\bar{g}_{33}}\left(\frac{\partial \bar{g}_{33}}{\partial \bar{x}^2}\right)\right] = 0$$

Using the fact that $\bar{g}/\bar{g}_{33} = (\bar{g}_{11}\bar{g}_{22} - \bar{g}_{12}\bar{g}_{21})$, the partial derivatives can be combined into one, as follows:

$$\frac{\partial \bar{v}^2}{\partial \bar{x}^2} + \frac{\bar{v}^2}{2\bar{g}}\frac{\partial \left[\bar{g}_{11}\bar{g}_{22} - \bar{g}_{12}\bar{g}_{12}\ \bar{g}_{33}\right]}{\partial \bar{x}^2} = 0. \tag{15}$$

The quantity in brackets is recognized as the determinant, $\bar{g}$; therefore the continuity equation reduces to:

$$\frac{\partial \left[\bar{g}\bar{v}^2\bar{v}^2\right]}{\partial \bar{x}^2} = 0. \tag{16}$$

Integrating with respect to $\bar{x}^2$ gives:

$$\bar{g}\bar{v}^2\bar{v}^2 = F(\bar{x}^1)$$

In general, the constant of integration, $F$, may depend on $\bar{x}^1$. Solving for velocity gives:

$$\bar{v}^2 = \frac{f(\bar{x}^1)}{\sqrt{\bar{g}}}, \qquad \bar{v}^1, \bar{v}^3 = 0 \tag{17}$$

Therefore, in the toroidal coordinate system, the velocity along any given streamline is inversely proportional to the Jacobian determinant. This reciprocal dependence on the Jacobian also holds for planar elliptical and circular vortices regardless of the vorticity distribution. At this point, specifics about the coordinate system have not been introduced. In obtaining (17), the only restrictions employed were that contours of one of the coordinates coincide with streamlines of the flow and that the velocity in the streamlined coordinate system is independent of the third coordinate direction. These two conditions can be restated as follows:

1. The coordinate system must be such that all quantities are independent of one of the coordinate directions in a subspace $V$ of $\mathbb{R}^3$ (in our case, $\bar{x}^3 \in V \subset \mathbb{R}^3$),

2. the velocity components in the remaining quotient space $M^2 = \mathbb{R}^3/V$ are parameterized by a single coordinate ($\bar{x}^2$ in our case); the other coordinate in $M^2$ ($\bar{x}^1$ in our case) is a constant of the motion.

Computing the vorticity using (17) shows that $\bar{\omega}^1, \bar{\omega}^2 = 0$ in the toroidal coordinate system. For the remaining coordinate direction, it will be useful to define the average of the vorticity tensor as follows:

$$\bar{\omega}_A^3 = \frac{\int \bar{\omega}^3 \mathrm{d}\bar{S}_3}{\int \mathrm{d}\bar{S}_3} \tag{18}$$

This allows the integral for the circulation to be performed on known geometrical parameters, as follows:

$$\Gamma = \int_{\bar{S}} \bar{\omega}^3 \mathrm{d}\bar{S}_3 = \bar{\omega}_A^3 \int_{\bar{S}} \mathrm{d}\bar{S}_3 \tag{19}$$

Note that the denominator of (18) is not the physical area but rather

$$\bar{S}_3(\bar{x}^1) \equiv \bar{S}_3 = \int_{\bar{S}} \mathrm{d}\bar{S}_3 = \int_0^{2\pi} \int_0^{\bar{x}^1} \sqrt{\bar{g}}\, \mathrm{d}\bar{x}^1 \mathrm{d}\bar{x}^2 . \tag{20}$$

The physical area is:

$$\bar{S}(3) = \int_{\bar{S}} \sqrt{\mathrm{d}\bar{S}^3 \mathrm{d}\bar{S}_3} = \int_{\bar{S}} \sqrt{\bar{g}^{33} \mathrm{d}\bar{S}_3 \mathrm{d}\bar{S}_3} = \int_{\bar{S}} \sqrt{\bar{g}^{33}} \mathrm{d}\bar{S}_3 = \int_{\bar{S}} \frac{\mathrm{d}\bar{S}_3}{\sqrt{\bar{g}_{33}}}$$

$$= \int_0^{2\pi} \int_0^{\bar{x}^1} \frac{\sqrt{\bar{g}}}{\sqrt{\bar{g}_{33}}} \mathrm{d}\bar{x}^1 \mathrm{d}\bar{x}^2$$

The formula for the circulation will be used to obtain an expression for the velocity field. We begin by writing the area integral:

$$\Gamma = \int_S \nabla \times \boldsymbol{v} \cdot \mathrm{d}\boldsymbol{S} = \int_{\bar{S}} \frac{\varepsilon^{ijk}}{\sqrt{\bar{g}}} \bar{v}_{k,j} \mathrm{d}\bar{S}_i \tag{21}$$

We raise the index of the velocity tensor in (21) and let $i = 3$ since $\bar{\omega}^3$ is the only nonzero vorticity component. Then since $\bar{v}^2$ is the only nonzero velocity component in the toroidal coordinate system, we have

$$\Gamma = \int_0^{2\pi} \int_0^{\bar{x}^1_{\max}} \left[ \frac{\partial\, \bar{g}_{22} \bar{v}^2}{\partial \bar{x}^1} - \frac{\partial\, \bar{g}_{12} \bar{v}^2}{\partial \bar{x}^2} \right] \mathrm{d}\bar{x}^1 \mathrm{d}\bar{x}^2 . \tag{22}$$

Note that both terms in the integrand are nonzero locally and therefore contribute to the local vorticity tensor (see (53)). However, the second term integrates to zero so that the circulation can be calculated by the first term alone. Continuing with the integration of the first term in (22) gives:

$$\Gamma = \int_0^{2\pi} \bar{g}_{22} \bar{v}^2 \Big|_{\bar{x}^1_{\max}} \mathrm{d}\bar{x}^2 - \int_0^{2\pi} \bar{g}_{22} \bar{v}^2 \Big|_{\bar{x}^1=0} \mathrm{d}\bar{x}^2 . \tag{23}$$

By evaluating $\bar{g}_{22}$ at $\bar{x}^1 = 0$, we find that the far right term of (23) is zero, and we are left with a single closed line integral along any streamline (in this case the streamline corresponding to $\bar{x}^1_{\max}$). This proves Stokes' theorem for the present coordinate system, namely:

$$\int \boldsymbol{\omega} \cdot \mathrm{d}\boldsymbol{S} = \oint \boldsymbol{v} \cdot \mathrm{d}\boldsymbol{s}$$

or in index notation:

$$\int_S \bar{\omega}^k \, \mathrm{d}\bar{S}_k = \oint \bar{g}_{ij} \bar{v}^i \, \mathrm{d}\bar{x}^j, \tag{24}$$

where the closed line integral along a curve of constant $\bar{x}^1$ encloses the area $S$. Since $\bar{v}^2$ is the only nonzero component in the toroidal coordinate system, $i=2$ in (24). The integral of the velocity is performed over the coordinate $\bar{x}^2$ with $\bar{x}^1$ and $\bar{x}^3$ held constant; therefore, $j=2$. Applying Stokes' theorem on any streamline therefore gives:

$$\Gamma(\bar{x}^1) = \oint \bar{g}_{22} \bar{v}^2 \Big|_{\bar{x}^1} \mathrm{d}\bar{x}^2, \tag{25}$$

where $\bar{v}^2$ is given by (17). Recall that since the function $f$ in (17) is dependent only upon $\bar{x}^1$, it remains constant along any given streamline. This is important because it allows this unknown function $f$ to be pulled out of the integral in (25), as follows:

$$\Gamma(\bar{x}^1) = f(\bar{x}^1) \oint \left( \frac{\bar{g}_{22}}{\sqrt{\bar{g}}} \right) \Big|_{\bar{x}^1} \mathrm{d}\bar{x}^2. \tag{26}$$

Therefore, the integral for the circulation can be performed on known properties of the coordinate system.

It is at this point that details about the coordinate system metric tensor must be utilized in (26). Doing so gives:

$$\Gamma(\bar{x}^1) = f(\bar{x}^1) \bar{x}^1 \bar{x}^1 \int_0^{2\pi} \frac{B^2 + \sin^2(\bar{x}^2)}{\sqrt{\bar{g}}} \mathrm{d}\bar{x}^2$$

$$\Gamma(\bar{x}^1) = \frac{\bar{x}^1 f(\bar{x}^1)}{AB} \int_0^{2\pi} \frac{B^2 + \sin^2(\bar{x}^2)}{[\bar{x}^1 B(\sin(\bar{x}^2) - k) + R_c][1 - k\sin(\bar{x}^2)]} \mathrm{d}\bar{x}^2$$

The integral can be simplified somewhat by first performing the polynomial division in the integrand, to give:

$$\Gamma(\bar{x}^1) = \frac{f(\bar{x}^1)}{AB^2} \left[ \int_0^{2\pi} \frac{1}{-k} \mathrm{d}\bar{x}^2 + \int_0^{2\pi} \frac{\alpha \sin(\bar{x}^2) + \beta}{[\sin(\bar{x}^2) + c][1 - k\sin(\bar{x}^2)]} \mathrm{d}\bar{x}^2 \right] \tag{27}$$

where

$$\alpha = \alpha(\bar{x}^1) = \left( \frac{1}{k} - c \right), \qquad \beta = \beta(\bar{x}^1) = \left( B^2 + \frac{c}{k} \right),$$

and

$$c = c(\bar{x}^1) = \frac{R_c}{\bar{x}^1 B} - k \tag{28}$$

Using partial fractions of the second term in (27), the integral can be simplified still further:

$$\Gamma(\bar{x}^1) = \frac{f(\bar{x}^1)}{AB^2}\left[\int_0^{2\pi}\frac{1}{-k}\,d\bar{x}^2 + \int_0^{2\pi}\frac{-\alpha c + \beta}{[1+kc][\sin(\bar{x}^2)+c]}\,d\bar{x}^2 + \int_0^{2\pi}\frac{\alpha + \beta k}{[1+kc][1-k\sin(\bar{x}^2)]}\,d\bar{x}^2\right]$$

Integrating then gives:

$$\Gamma(\bar{x}^1) = \frac{f(\bar{x}^1)}{AB^2}\left[2\pi\left(\frac{1}{-k} + \frac{-\alpha c + \beta}{(1+kc)\sqrt{c^2-1}} + \frac{\alpha + \beta k}{(1+kc)\sqrt{1-k^2}}\right)\right]$$

which can be further simplified to:

$$\Gamma(\bar{x}^1) = \frac{f(\bar{x}^1)}{AB^2}\left[2\pi\left(\frac{1}{-k} + \frac{B^2 + c^2}{(1+kc)\sqrt{c^2-1}} + \frac{1/k + B^2 k}{(1+kc)\sqrt{1-k^2}}\right)\right]$$

or

$$\Gamma(\bar{x}^1) = \frac{2\pi f(\bar{x}^1)}{AB^2}\left[P(\bar{x}^1) + Q(\bar{x}^1)\right], \tag{29}$$

where

$$P(\bar{x}^1) = \frac{B^2 + c^2}{(1+kc)\sqrt{c^2-1}} \quad \text{and} \quad Q(\bar{x}^1) = \frac{(-1/k - c)\sqrt{1-k^2} + 1/k + B^2 k}{(1+kc)\sqrt{1-k^2}}.$$

Note that $Q(\bar{x}^1) \to -c(\bar{x}^1)$ as the offset $k \to 0$. The circulation, as written in (29), is a function of only $\bar{x}^1$. Moreover, it is known from Eq. (19) that this dependence is purely geometrical; that is,

$$\Gamma(\bar{x}^1) = \bar{\omega}_A^3 \bar{S}_3(\bar{x}^1). \tag{30}$$

To transform (29) into an equation with no dependence on $\bar{x}^1$, we use Eq. (30) to construct the requirement that the ratio $\Gamma(\bar{x}^1) / \bar{S}_3(\bar{x}^1)$ evaluated at, say, the outer streamline ($\bar{x}^1 = \bar{x}^1_{\max}$) be equal to the same quantity evaluated along any other streamline. In other words, we want: $\bar{\omega}_A^3(\bar{x}^1) = \bar{\omega}_A^3(\bar{x}^1_{\max})$ for all $\bar{x}^1$ in the vortex. We therefore set

$$1 = \frac{\bar{\omega}_A^3(\bar{x}^1_{\max})}{\bar{\omega}_A^3(\bar{x}^1)} = \frac{\bar{S}_3(\bar{x}^1)\,f(\bar{x}^1_{\max})}{\bar{S}_3(\bar{x}^1_{\max})f(\bar{x}^1)}\frac{[P(\bar{x}^1_{\max}) + Q(\bar{x}^1_{\max})]}{[P(\bar{x}^1) + Q(\bar{x}^1)]} \tag{31}$$

The function $f(\bar{x}^1_{\max})$ corresponding to the outermost streamline will be eliminated by expressing it in terms of the physical velocity, $v_O$, at the point on the perimeter of the toroidal vortex farthest from the axis of symmetry, which is taken as known. This is done by first finding the expression for the magnitude of the velocity, as follows:

$$|\mathbf{v}|^2 = \bar{g}_{ij}\bar{v}^i\bar{v}^j$$

$$\therefore |\mathbf{v}| = \sqrt{\bar{g}_{22}}\,\bar{v}^2$$

$$|\mathbf{v}| = \bar{x}^1\sqrt{B^2 + \sin^2(\bar{x}^2)}\,\bar{v}^2 \tag{32}$$

Substituting (17) into (32) and evaluating at the point $(\bar{x}^1, \bar{x}^2) = (\bar{x}^1_{\max}, \pi/2)$ gives:

$$v_O = \frac{\bar{x}^1_{\max}\sqrt{B^2+1}\,f(\bar{x}^1_{\max})}{\sqrt{g(\bar{x}^1_{\max},\pi/2)}}$$

where $v_O$ is the velocity at the outer perimeter point and, from (9),

$$\sqrt{g(\bar{x}^1_{\max},\pi/2)} = R_O\,\bar{x}^1_{\max}AB(1-k)\,.$$

Therefore,

$$v_O = \frac{f(\bar{x}^1_{\max})}{R_O B(1-k)}\,. \tag{33}$$

Here, the useful identity in (6) has been used. Solving (31) for $f(\bar{x}^1)$ and using (33) to eliminate $f(\bar{x}^1_{\max})$ from the result gives:

$$f(\bar{x}^1) = v_O R_O B(1-k)\frac{\bar{S}_3(\bar{x}^1)}{\bar{S}_3(\bar{x}^1_{\max})}\frac{[P(\bar{x}^1_{\max}) + Q(\bar{x}^1_{\max})]}{[P(\bar{x}^1) + Q(\bar{x}^1)]}$$

$$f(\bar{x}^1) = v_O R_O B(1-k)\left(\frac{\bar{x}^1}{\bar{x}^1_{\max}}\right)^3\frac{c}{c_{\max}}\frac{[P(\bar{x}^1_{\max}) + Q(\bar{x}^1_{\max})]}{[P(\bar{x}^1) + Q(\bar{x}^1)]}\,, \tag{34}$$

where the relation $\bar{S}_3 = \pi AB^2(\bar{x}^1)^3 c$ obtained from (20) has been used to eliminate $\bar{S}_3(\bar{x}^1)$. The simplest way to determine the average of the vorticity tensor is to write (29) and (30) for $\bar{x}^1_{\max}$ rather than $\bar{x}^1$, eliminate $\Gamma(\bar{x}^1_{\max})$ from the two equations, and solve for $\bar{\omega}^3_A$. Equation (33) is then substituted into the result, to give:

$$\bar{\omega}^3_A = \frac{8 v_O R_O(R_O - R_C)[P(\bar{x}^1_{\max}) + Q(\bar{x}^1_{\max})]}{a^2 B^2 (R_O^2 - R_I^2)}\,. \tag{35}$$

Here, the fact that $(1-k)/c_{\max} = 2(R_O - R_C)/(R_O + R_I)$ [Eqs. (28) and (5)] has been used. As indicated by Eq. (35), the average of the vorticity tensor is uniform, being independent of $\bar{x}^1$, $\bar{x}^2$, and $\bar{x}^3$.

An expression for the velocity in the toroidal coordinates can be obtained by substituting Eq. (34) into Eq. (17). The result is:

$$\bar{v}^2 = \frac{2 v_O R_O B\,c}{\sqrt{g}}\left(\frac{\bar{x}^1}{\bar{x}^1_{\max}}\right)^3\frac{(R_O - R_C)}{R_O + R_I}\frac{[P(\bar{x}^1_{\max}) + Q(\bar{x}^1_{\max})]}{[P(\bar{x}^1) + Q(\bar{x}^1)]} \tag{36}$$

This solution can be verified by substituting (35) and (36) into (24).

### 2.3. *Rectangular Coordinates*

The velocity distribution can be transformed to a rectangular coordinate system by means of the transformation $v^i = (\partial x^i/\partial \bar{x}^j)\bar{v}^j$, which gives:

$$v^i = \begin{bmatrix} \cos(\hat{x}^3)\bar{x}^1 B\cos(\bar{x}^2) \\ \sin(\hat{x}^3)\bar{x}^1 B\cos(\bar{x}^2) \\ -\bar{x}^1 A\sin(\bar{x}^2) \end{bmatrix}\bar{v}^2\,. \tag{37}$$

The velocity field can be viewed on any plane through the axis of symmetry. For example, on the plane made by, say, $\hat{x}^3 = \pi/2$, we have,

$$v^i = \begin{bmatrix} 0 \\ \bar{x}^1 B \cos(\bar{x}^2) \\ -\bar{x}^1 A \sin(\bar{x}^2) \end{bmatrix} \bar{v}^2 \tag{38}$$

### 2.4. *The Stream Function*

The purpose of the stream function is to map two independent variables into a single dependent variable, $\psi$. In the present work, this is accomplished with the variable $\bar{v}^2$, whose connection with the stream function will now be explained.

The stream function is found by choosing the function so that the continuity equation is satisfied identically due to a cancellation of identical terms. The stream function for two-dimensional flow has been found to be related to components of the velocity tensor by:

$$v^\alpha = \frac{J^{\alpha\beta}}{\sqrt{g}} \frac{\partial \psi}{\partial x^\beta} \qquad (\alpha,\ \beta = 1,\ 2) \tag{39}$$

where $\sqrt{g}$ is the Jacobian of the pertinent coordinate system relative to a Cartesian system, and $J^{\alpha\beta}$ is the symplectic matrix given by:

$$J^{\alpha\beta} = \begin{bmatrix} 0 & 1 \\ -1 & 0 \end{bmatrix}.$$

Equation (39) holds for stream functions in rectangular, cylindrical, and spherical coordinate systems, as well as the toroidal coordinate system studied here. In spherical coordinates, for example, (39) results in the Stokes stream function. By comparing (39) with (17), we see that

$$\frac{\partial \psi}{\partial \bar{x}^1} = -f(\bar{x}^1). \tag{40}$$

This equation does, in fact, force the continuity equation (16), which can be restated as

$$\frac{\partial \sqrt{g} \bar{v}^2}{\partial \bar{x}^2} = 0, \tag{41}$$

to be satisfied identically because the choice of the coordinate system renders the stream function independent of $\bar{x}^2$. The coordinate $\bar{x}^2$ is an angle variable, and $\bar{x}^1$ plays the role of an action variable (Arnold 1978; Fountain *et al.* 2000). Equation (41) is valid regardless of the vorticity distribution. To see that (41) is satisfied by (39), substitute (39) into (41), giving

$$\frac{\partial}{\partial \bar{x}^2}\left[\sqrt{g}\left(\frac{-1}{\sqrt{g}}\frac{\partial \psi}{\partial \bar{x}^1}\right)\right] = -\frac{\partial}{\partial \bar{x}^2}\left(\frac{\partial \psi}{\partial \bar{x}^1}\right)$$

which, by (40), is zero. Consequently, the stream function can be found by integrating

(40), as follows:
$$\psi = -\int f(\bar{x}^1)\, d\bar{x}^1. \tag{42}$$

The following discussion is not restricted to the toroidal coordinate system and vortex studied thus far. Therefore, quantities in such general discussions will not be written with overbars. By reasoning similar to that used to derive (15), the general continuity equation for the steady flow of a variable-density fluid, with $v^1$, $v^2$, and $v^3$ nonzero, can be shown to be:

$$\rho \frac{\partial v^i}{\partial x^i} + \frac{\rho v^i}{2g} \frac{\partial g}{\partial x^i} + v^i \frac{\partial \rho}{\partial x^i} = 0. \tag{43}$$

Grouping terms as done in obtaining (16) reduces the continuity equation to:

$$\frac{1}{\rho\sqrt{g}} \frac{\partial}{\partial x^i} (\rho\sqrt{g}\, v^i) = 0 \tag{44}$$

The continuity equation given in (44), which states that the Lie derivative of the quantity $\rho\sqrt{g}$ is zero, is valid for any steady flow in any coordinate system. For complicated coordinate systems, this form is more convenient than formulations such as (12). For an incompressible fluid the density, $\rho$, in (44) can be omitted. If the flow is such that quantities do not depend upon, say, $x^3$ (condition 1 of §2.2), we can write (44) as two separate statements:

$$\frac{\partial}{\partial x^\alpha} (\rho\sqrt{g}\, v^\alpha) = 0 \qquad (x^\alpha \in M^2) \tag{45}$$

and

$$\frac{\partial}{\partial x^3} (\rho\sqrt{g}\, v^3) = 0. \qquad (x^3 \in V) \tag{46}$$

where $M^2 = \mathbb{R}^3 / V$. Equation (45) leads immediately to a further generalization of Eq. (39) for compressible flows:

$$v^\alpha = \frac{J^{\alpha\beta}}{\rho\sqrt{g}} \frac{\partial \psi}{\partial x^\beta} \qquad (x^\alpha \in M^2) \tag{47}$$

where $\psi: M^2 \to \mathbb{R}$ is the stream function for any coordinate system parametrizing $M^2$. Substituting (47) into (45) and expanding, we find that:

$$\frac{\partial}{\partial x^1}\left(\frac{\partial \psi}{\partial x^2}\right) - \frac{\partial}{\partial x^2}\left(\frac{\partial \psi}{\partial x^1}\right) = 0,$$

which shows the cancellation of identical terms typically seen in stream function formulations. The difference here is that the space $M^2$ spanned by $x^1$ and $x^2$ need not be planar, and the coordinate system need not be linear nor orthogonal. The velocity component in $V$ need not be zero but must satisfy (46). Equations (45) and (46) and the stream function in (47) are valid for any flow for which some subspace $V$, sufficient to satisfy (46), can be found and divided out of $\mathbb{R}^3$, leaving only the space $M^2$ in which to solve the two-dimensional portion of the problem. When streamline coordinates are found, condition 2 of §2.2 is also satisfied, and each term in (44) is zero individually, as in the case of the present toroidal vortex.

Using the Poisson bracket, given by

$$\{\cdot,\psi\} = \left(\frac{\partial \psi}{\partial x^2}\frac{\partial}{\partial x^1} - \frac{\partial \psi}{\partial x^1}\frac{\partial}{\partial x^2}\right),$$

(47) can also be written in the form:

$$\dot{\boldsymbol{x}} = \frac{1}{\rho\sqrt{g}}\{\boldsymbol{x},\psi\}$$

with $\boldsymbol{x} \in M^2$; $\psi : M^2 \to \mathbb{R}$.

### 2.5. Vorticity

It is well known that, in two-dimensional flow, replacing the velocity components in the vorticity definition ($\boldsymbol{\omega} = \nabla \times \boldsymbol{v}$) with derivatives of the stream function results in the following Poisson equation: $\omega = -\nabla^2\psi$. While this is true for rectangular and cylindrical coordinate systems, it is not true for the spherical coordinate system, nor is it true for the toroidal coordinate system used in this work. The more general relation can be found by substituting (39) (or (47)) into the definition of vorticity written below:

$$\omega^i = E^{ijk}v_{k,j} = \frac{\varepsilon^{ijk}}{\sqrt{g}}\, g_{k\alpha}\, v^\alpha_{\ ,j}$$

The substitution gives

$$\omega^i = \frac{\varepsilon^{ijk}}{\sqrt{g}}\frac{\partial}{\partial x^j}\left(\frac{g_{k\alpha}J^{\alpha\beta}}{\sqrt{g}}\frac{\partial \psi}{\partial x^\beta}\right). \qquad (\alpha, \beta \neq i) \qquad (48)$$

To find the vorticity in the third coordinate direction due to motion in the other two, we let $i = 3$ in (48). This gives:

$$\omega^3 = \frac{1}{\sqrt{g}}\left[\frac{\partial}{\partial x^1}\left(\frac{g_{2\alpha}J^{\alpha\beta}}{\sqrt{g}}\frac{\partial \psi}{\partial x^\beta}\right) - \frac{\partial}{\partial \overline{x}^2}\left(\frac{g_{1\alpha}J^{\alpha\beta}}{\sqrt{g}}\frac{\partial \psi}{\partial x^\beta}\right)\right] \qquad (49)$$

Expanding the indices and noting that $J^{\alpha\beta} = 0$ when $\alpha = \beta$, gives:

$$\omega^3 = \frac{1}{\sqrt{g}}\left[\frac{\partial}{\partial x^1}\left(\frac{g_{21}}{\sqrt{g}}\frac{\partial \psi}{\partial x^2}\right) - \frac{\partial}{\partial \overline{x}^1}\left(\frac{g_{22}}{\sqrt{g}}\frac{\partial \psi}{\partial x^1}\right) - \frac{\partial}{\partial x^2}\left(\frac{g_{11}}{\sqrt{g}}\frac{\partial \psi}{\partial x^2}\right) + \frac{\partial}{\partial x^2}\left(\frac{g_{12}}{\sqrt{g}}\frac{\partial \psi}{\partial x^1}\right)\right] \qquad (50)$$

In rectangular and cylindrical coordinate systems, the right side of (50) reduces to the two-dimensional Laplacian operator. For example, in the cylindrical coordinate system (with $\hat{g}_{22} = R^2$) (50) becomes:

$$\hat{\omega}^3 = \frac{-1}{\sqrt{\hat{g}}}\frac{\partial}{\partial \hat{x}^1}\left(\frac{\hat{g}_{22}}{\sqrt{\hat{g}}}\frac{\partial \psi}{\partial \hat{x}^1}\right) - \frac{1}{\sqrt{\hat{g}}}\frac{\partial}{\partial \hat{x}^2}\left(\frac{\hat{g}_{11}}{\sqrt{\hat{g}}}\frac{\partial \psi}{\partial \hat{x}^2}\right) \qquad (51)$$

or

$$-\omega = \frac{1}{R}\frac{\partial}{\partial R}\left(R\frac{\partial \psi}{\partial R}\right) + \frac{1}{R^2}\frac{\partial^2 \psi}{\partial \theta^2}, \qquad (52)$$

which is indeed the Poisson equation, $\nabla^2\psi = -\omega$, in cylindrical coordinates. When $\omega = 0$ or $\omega = -\lambda^2\psi$, (52) is separable by assuming a product solution, the case $\omega = -\lambda^2\psi$ leading to a Bessel equation.

Returning now to the more general (48), if we require that the coordinate system coincide with streamline coordinates, then we obtain (49) for the 3$^{\text{rd}}$ coordinate direction. And since $\partial \psi / \partial \bar{x}^2 = 0$, we can make the limitation $\beta = 1$ so that $\alpha = 2$. The result is:

$$\bar{\omega}^3 = \frac{-1}{\sqrt{\bar{g}}}\left[\frac{\partial}{\partial \bar{x}^1}\left(\frac{\bar{g}_{22}}{\sqrt{\bar{g}}}\frac{d\psi}{d\bar{x}^1}\right) - \frac{\partial}{\partial \bar{x}^2}\left(\frac{\bar{g}_{12}}{\sqrt{\bar{g}}}\frac{d\psi}{d\bar{x}^1}\right)\right], \tag{53}$$

which is valid for steady vortex flows satisfying conditions 1 and 2 of §2.2, regardless of the vorticity distribution. Note that when calculating the local vorticity tensor, the second term on the right side of (53) is nonzero. Recall, however, that because of its symmetry properties, it integrates to zero when calculating the circulation (22). To see this, substitute (53) back into (19) and perform the integration.

Equation (50) is not separable; however, by the change to streamline coordinates, (53) is separable. This is because $\partial \psi / \partial \bar{x}^2 = 0$. Rearranging (53) as

$$\bar{\omega}^3 = -\frac{\bar{g}_{22}}{\bar{g}}\frac{d^2\psi}{(d\bar{x}^1)^2} + \frac{1}{\sqrt{\bar{g}}}\left[\frac{\partial}{\partial \bar{x}^2}\left(\frac{\bar{g}_{12}}{\sqrt{\bar{g}}}\right) - \frac{\partial}{\partial \bar{x}^1}\left(\frac{\bar{g}_{22}}{\sqrt{\bar{g}}}\right)\right]\frac{d\psi}{d\bar{x}^1} \tag{54}$$

reveals the form $\psi'' + P(\bar{x}^1)\psi' = h(\bar{x}^1)$, where the functions $P$ and $h$ depend on $\bar{x}^2$ as well as $\bar{x}^1$.

Summarizing, in streamline coordinates the general form of the function $\mathcal{L}$ in the relation

$$\omega = \mathcal{L}(\psi) \tag{55}$$

is:

$$\mathcal{L} = -\frac{\bar{g}_{22}}{\bar{g}}\frac{d^2}{(d\bar{x}^1)^2} + \frac{1}{\sqrt{\bar{g}}}\left[\frac{\partial}{\partial \bar{x}^2}\left(\frac{\bar{g}_{12}}{\sqrt{\bar{g}}}\right) - \frac{\partial}{\partial \bar{x}^1}\left(\frac{\bar{g}_{22}}{\sqrt{\bar{g}}}\right)\right]\frac{d}{d\bar{x}^1} \tag{56}$$

If a streamline coordinate system cannot be found, then Eq. (48) must be used in place of (54). By referring to (50), (48) can be cast in terms of the symplectic matrix $J^{\alpha\beta}$, as follows:

$$\omega^i = \frac{J^{\gamma\eta}J^{\alpha\beta}}{\sqrt{g}}\frac{\partial}{\partial x^\alpha}\left(\frac{g_{\beta\gamma}}{\sqrt{g}}\frac{\partial \psi}{\partial x^\eta}\right) \qquad (\alpha,\beta,\gamma,\eta \neq i) \tag{57}$$

In complex-lamellar flows, wherein $\boldsymbol{\omega} \cdot \boldsymbol{v} = 0$, coordinate systems in which $v^i = v^i(x^j)$ with $x^j \in \mathbb{R}^3$ can be transformed to one wherein $\bar{v}^i = \bar{v}^i(\bar{x}^\alpha)$ with $\bar{x}^\alpha \in M^2 = \mathbb{R}^3/V$ and $\bar{\omega}^3 \in V$. In our streamline coordinate system, condition 2 in §2.2 is also satisfied so that (55) is separable. This is because, within the two-dimensional manifold $M^2$, contours of $\bar{x}^2$ follow an invariant set that is selected by changing $\bar{x}^1$.

### 2.6. *Angular Momentum*

The angular momentum within the vortex is given by:

$$\bar{p}_i = \left.\boldsymbol{x} \times \rho \boldsymbol{v}\right._i = \varepsilon_{ijk}\rho\sqrt{\bar{g}}\,\bar{x}^j\,\bar{v}^k.$$

Using (47) to represent $\bar{v}^k$, we have:

$$\tag{58}$$

$$\overline{p}_i = -\varepsilon_{ijk}\,\overline{x}^j\,J^{kp}\,\frac{\partial \psi}{\partial \overline{x}^p}.$$

Since $J^{kp} = 0$ when $k = p$, (58) reduces to

$$\overline{p}_i = -\overline{x}^\alpha \frac{\partial \psi}{\partial \overline{x}^\alpha} \qquad (\alpha \neq i) \qquad (59)$$

If $\overline{p}_i \in V \subset \mathbb{R}^3$, then $\alpha$ is summed over the two-dimensional quotient space $M^2 = \mathbb{R}^3 / V$. For complex-lamellar flows, $V$ is the one-dimensional subspace containing the vortex axis of rotation. For the problem at hand, (59) is applied in streamline coordinates with $\overline{p}_3 \in V$; therefore, $i = 3$ in (59):

$$\overline{p}_3 = -\overline{x}^1 \frac{d\psi}{d\overline{x}^1} \qquad (60)$$

Therefore, the angular momentum throughout the core of the toroidal vortex is independent of $\overline{x}^2$ and $\overline{x}^3$, remaining constant during the motion.

## 3. Results and Discussion

Since $\overline{x}^3$ and $\widetilde{x}^3$ are coincident (compare Figure 1 and Figure 2), the vorticity tensors in the spherical and toroidal coordinate systems are both directed along the same curves. In fact, by the same reasoning used to obtain (3), the physical component of vorticity in the core of the vortex ring is:

$$\overline{\omega}(3) = \sqrt{g_{33}}\,\overline{\omega}^3 = R\,\overline{\omega}^3$$

which is the same relation found for Hill's spherical vortex. The quantity $R$ in both cases is the distance from the axis of symmetry.

The cross section area of the core of Hill's spherical vortex differs from the cross section area of the sphere itself by a factor of 2. Therefore, the ratio of the radius of the sphere to the mean core radius of the spherical vortex is $\sqrt{2}$. For a vortex ring of elliptical cross section, the core area is $\pi ab$. Therefore, its mean core radius, $\xi$, is:

$$\xi = \sqrt{ab} = \sqrt{\left[\frac{a}{b}\right]b^2} = \frac{(R_O - R_I)}{2}\sqrt{\frac{a}{b}}$$

The parameter, $\alpha = \xi / R_C$, by which Norbury (1973) defined a family of vortex rings with constant $\omega(3)/R$ can be defined for vortex rings with elliptical cross sections as:

$$\alpha = \frac{(R_O - R_I)}{2R_C}\sqrt{\frac{a}{b}} \qquad (61)$$

where $R_O$, $R_C$, and $R_I$ are the distances from the axis of symmetry to the outer radius, the centre, and the inner radius, respectively. The latter also corresponds to the radius of the central jet penetrating the vortex ring. Gharib *et al.* (1998) show clearly this separation of the vortex region from the jet-like region along the axis of symmetry.

The velocity magnitude within the core of the vortex ring is found by substituting (36) for $\overline{v}^2$ into (32). Figures 3 through 5 show comparisons of the velocity profile through the vortex centre as a function of $R/R_C$, the dimensionless distance from the axis of symmetry. In all three cases, the shift constant, *k*, was chosen so as to eliminate any discontinuity in the vorticity profile. Figure 3 shows the velocity profile for the toroidal vortex ring and for Hill's spherical vortex having equal outer radii, $R_O$, outer velocities, $v_O$, and circulation. Unlike Hill's spherical vortex, however, this requires that $\alpha < \sqrt{2}$ and $v_I/v_O > 1$ for the vortex ring. This is due to the topological difference between the vortex ring and the spherical vortex.

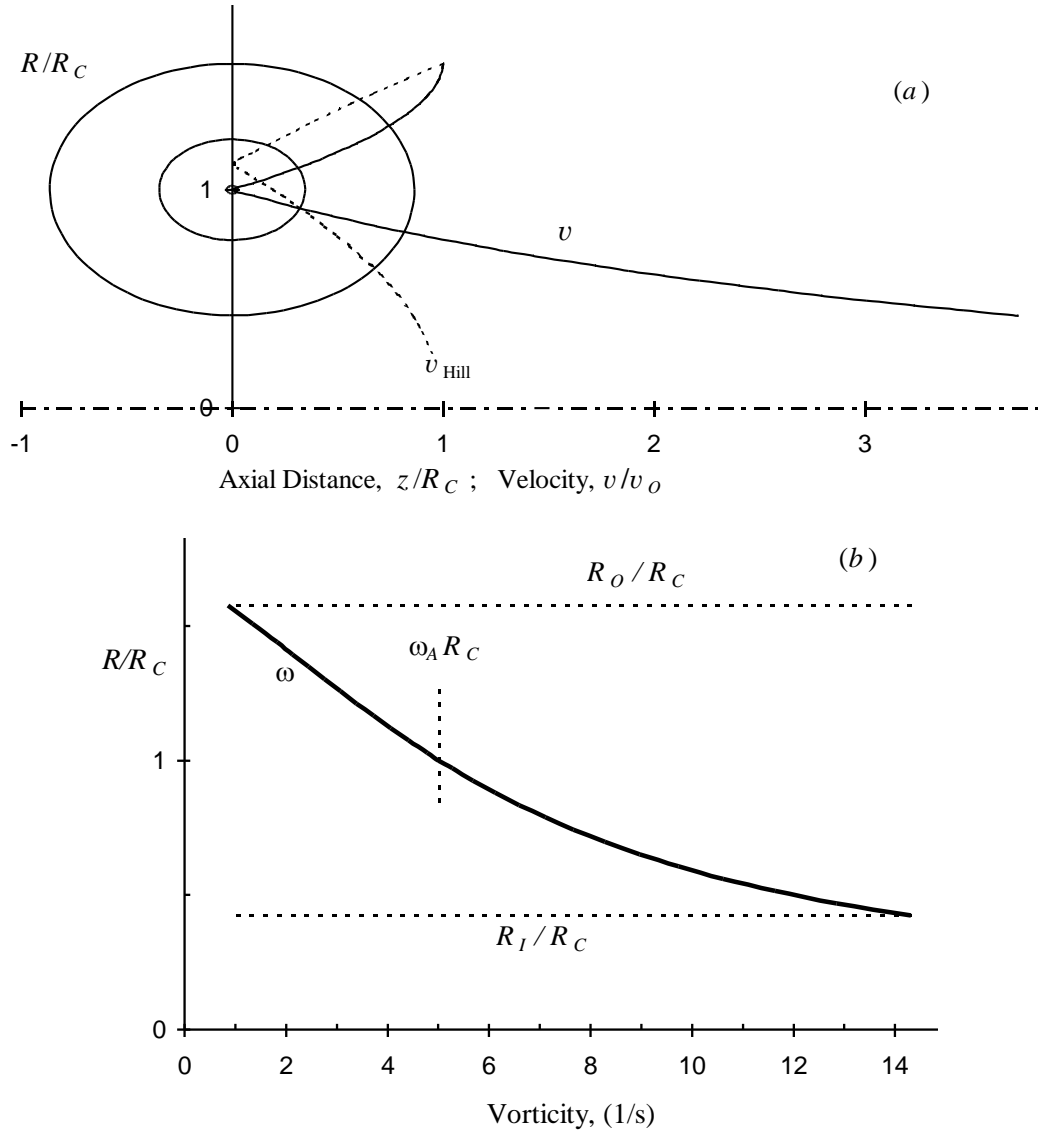

Figure 3. Toroidal vortex with $v_O = 1$, axis ratio $a/b = 1.5$, $R_O/R_C = 1.576$, $\alpha = 0.706$, offset $k = 0.0$, and $\Gamma/\Gamma_{\text{Hill}} = 1.0$. (*a*) Velocity magnitude through the vortex centre for the present vortex ring (solid curve) and for Hill's spherical vortex (dashed curve), (*b*) physical vorticity profile through centre.

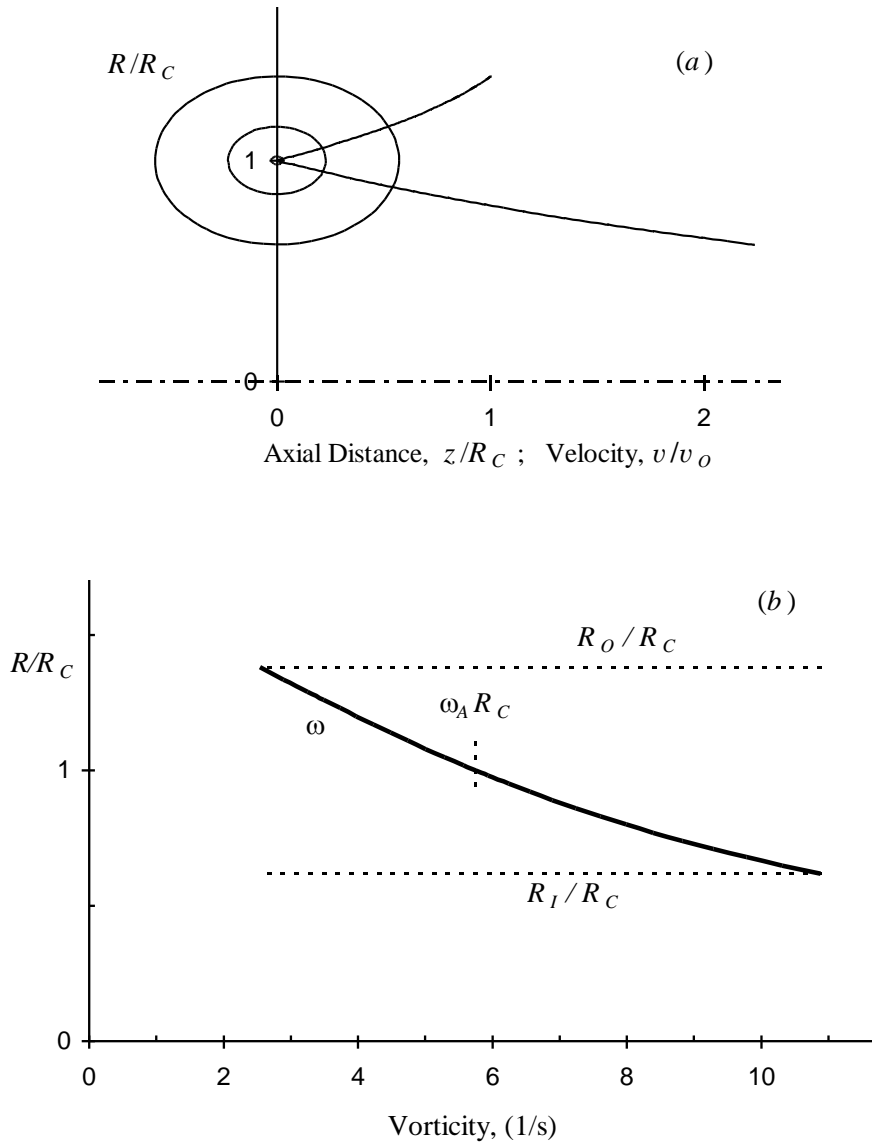

Figure 4. Toroidal vortex with $v_O = 1.0$, axis ratio $a/b = 1.5$, $R_O/R_C = 1.381$, $\alpha = 0.466$, offset $k = 0.0$, and $\Gamma/\Gamma_{\text{Hill}} = 0.5$. (a) Velocity magnitude through the vortex centre, (b) physical vorticity profile through centre.

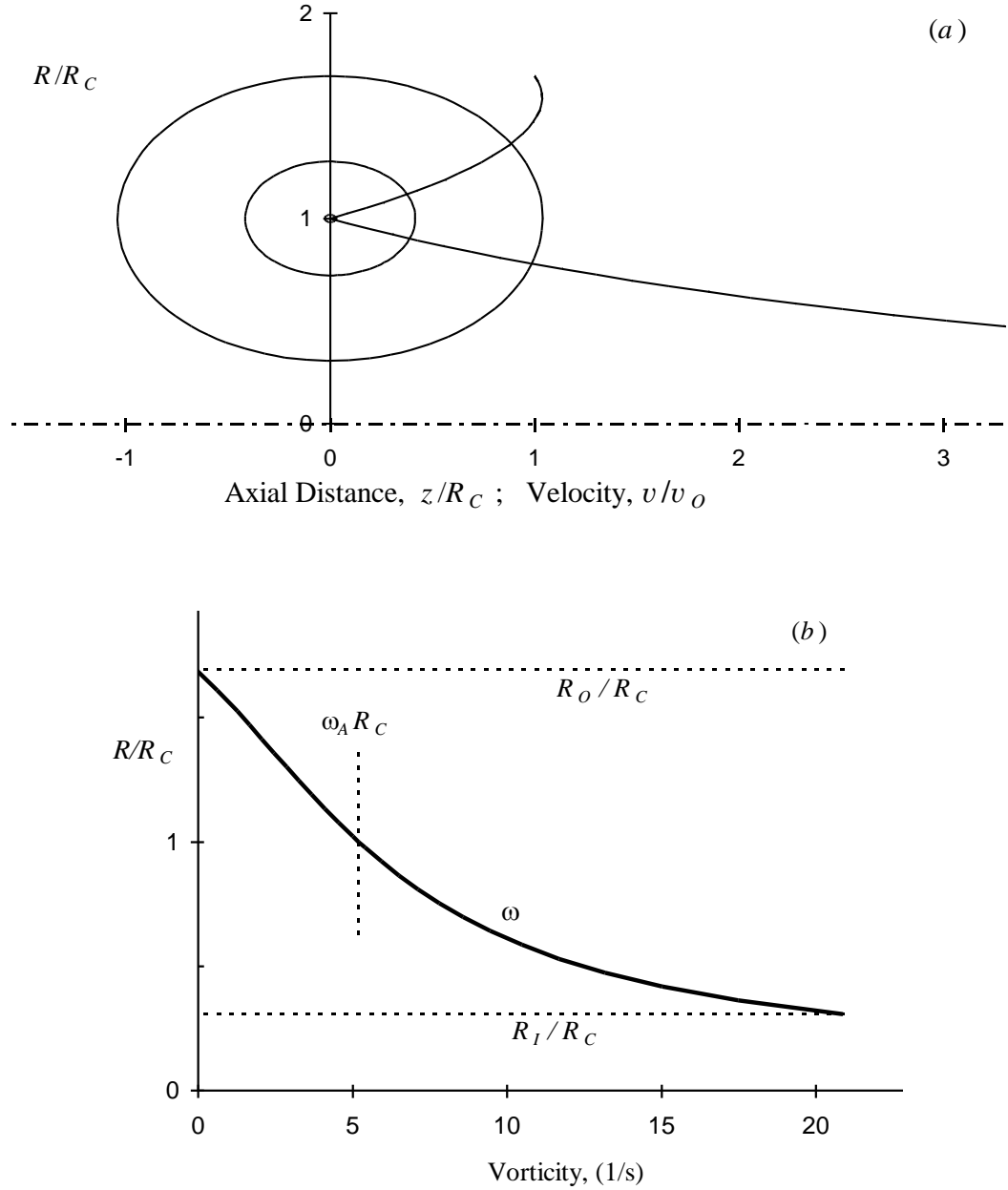

Figure 5. Toroidal vortex with $v_I/v_O = 5.523$, axis ratio $a/b = 1.5$, $R_O/R_C = 1.693$, $R_I/R_C = 0.307$, $\alpha = 0.849$, offset $k = 0.0$, and $\Gamma/\Gamma_{\text{Hill}} = 1.5$. (*a*) Velocity magnitude through the vortex centre, (*b*) physical vorticity profile through centre.

The possibility that the central jet velocity becomes large is purely the result of a constricted reverse flow area as $R_I \to 0$. Hill's spherical vortex also has an infinite increase in velocity along the axis of symmetry, but it is of a less apparent nature. The flow on the bounding streamline in the spherical vortex increases from a velocity of zero at the rear stagnation point to a finite value as it passes through the constriction during its reversal. In fact, the forward and rear stagnation points in Hill's spherical vortex are a direct result of the boundedness of the velocity that is assumed in obtaining its solution (see §1). The vortex ring does not have such a point of zero velocity anywhere on its boundary, and this allows it to attain large velocities as the reverse flow area decreases when $R_I$ becomes small. Note from Figure 4 and Figure 5 that for a given vortex outer radius, the circulation of the vortex ring can be either smaller or larger than that of Hill's spherical vortex.

As mentioned earlier, the velocities at the inner and outer radii on the vortex ring are equal in the case of Hill's spherical vortex. As seen from the above figures, this need not be true for a toroidal vortex ring. A simple relation can be given for this velocity ratio. From (32), the velocity magnitudes at the outer and inner radii of the toroidal vortex ring are, respectively:

$$v = a\,\bar{v}^2(\bar{x}^1_{\max}, \pm \pi/2) \tag{62}$$

From (36), the velocity at any point on the perimeter is, in the elliptical coordinate system,

$$\bar{v}_P^2 = \frac{v_O R_O B(1-k)}{\sqrt{g(\bar{x}^1_{\max},\theta)}}, \tag{63}$$

where the subscript "$P$" denotes "perimeter." To obtain the physical velocity magnitude at the outer and inner radii, substitute (63) into (62) and evaluate at the points $(\bar{x}^1_{\max}, \pm \pi/2)$, respectively. The result is:

$$\frac{v_I}{v_O} = \frac{R_O}{R_I}\frac{(1-k)}{(1+k)}$$

where $v_I$ is the velocity at the inner radius. Using (5) and the fact that $(R_O - R_I) = 2b$ gives:

$$\frac{v_I}{v_O} = \frac{R_O}{R_I}\frac{(R_O - R_C)}{(R_C - R_I)} \tag{64}$$

While the velocity field generally depends upon the axis ratio of the vortex ring, (64) indicates that the ratio $v_I/v_O$ does not. If fact, the effect of vortex axis ratio on the entire velocity profile along the centreline is minimal. If all streamlines are concentric with centres located by $k = 0$, then (64) simplifies to:

$$\frac{v_I}{v_O} = \frac{R_O}{R_I} \tag{65}$$

### 3.1. *Momentum*

The governing momentum equation is

$$\frac{1}{2}\rho \nabla \boldsymbol{v}\cdot\boldsymbol{v} \;-\; \rho\,\boldsymbol{v}\times\boldsymbol{\omega} \;=\; -\nabla p \;+\; \nu\nabla^2\boldsymbol{v}.$$

Shariff and Leonard (1992) noted that in steady planar vortices, nonlinear terms balance each other over a large portion of their cores, implying an inviscid vortex structure. They then pointed out, however, that for laminar vortex rings, it is difficult to determine the degree to which such balance is achieved. In order to study this question further, we write the momentum equation in an alternative form by taking its curl:

$$\boldsymbol{v}\cdot\nabla\boldsymbol{\omega} \;=\; \boldsymbol{\omega}\cdot\nabla\boldsymbol{v} \;+\; \nu\nabla^2\boldsymbol{\omega}. \tag{66}$$

In index notation, (66) is:

$$\bar{v}^k \bar{\omega}^i_{,k} \;=\; \bar{\omega}^j \bar{v}^i_{,j} \;+\; \nu \bar{g}^{pk} \bar{\omega}^i_{,kp}. \tag{67}$$

For the vortex ring studied herein, the middle term in (67), often referred to as the vortex stretching term, is zero. To see this, note that the only non-zero component of vorticity is $\bar{\omega}^3$; therefore, $j=3$ in (67). Since $\bar{v}^i$ is independent of $\bar{x}^3$, the term is zero. Therefore, the momentum equation reduces to:

$$\bar{v}^2 \bar{\omega}^3_{,2} \;=\; \nu \bar{g}^{pk} \bar{\omega}^3_{,kp}. \tag{68}$$

A solution for the case when the vorticity tensor is constant along any given streamline, as it is for steady planar inviscid vortex structures, would be very appealing since this would yield the simplification $\bar{\omega}^3_{,2}=0$ in (68). However, the monotonic decrease in the vorticity magnitude with distance from the symmetry axis, as depicted in Figures 3 through 5 (the vorticity tensor itself decreases even more sharply), indicates that, at least for the present vortex ring, $\bar{\omega}^3_{,2}\neq 0$. Therefore, the viscous term in (68) must also be non-zero.

The left side of (68) expands as follows:

$$\bar{v}^2 \bar{\omega}^3_{,2} \;=\; \bar{v}^2 \left(\frac{\partial \bar{\omega}^3}{\partial \bar{x}^2} + \bar{\Gamma}^3_{m2}\bar{\omega}^m\right).$$

The only nonzero component of the vorticity tensor is $\bar{\omega}^3$; therefore, $m=3$. Consequently,

$$\bar{\omega}^3_{,2} \;=\; \left(\frac{\partial \bar{\omega}^3}{\partial \bar{x}^2} + \frac{1}{2\bar{g}_{33}}\frac{\partial \bar{g}_{33}}{\partial \bar{x}^2}\bar{\omega}^3\right). \tag{69}$$

Note that the right side of (69) can be grouped as follows:

$$\bar{\omega}^3_{,2} \;=\; \frac{1}{\sqrt{\bar{g}_{33}}}\frac{\partial}{\partial \bar{x}^2}\sqrt{\bar{g}_{33}}\,\bar{\omega}^3. \tag{70}$$

This is simply the derivative of the physical component of vorticity in the 03 direction divided by $\sqrt{\bar{g}_{33}}$. Substituting (54) into (70) and consolidating in index notation gives:

$$\bar{\omega}^3_{,2} \;=\; \frac{J^{\alpha\beta}}{\sqrt{\bar{g}_{33}}}\frac{\partial}{\partial \bar{x}^2}\left[\frac{\sqrt{\bar{g}_{33}}}{\sqrt{\bar{g}}}\frac{\partial}{\partial \bar{x}^\beta}\left(\frac{\bar{g}_{\alpha 2}}{\sqrt{\bar{g}}}\psi'\right)\right]. \tag{71}$$

Note the relationship between (71) and (57), particularly when $\eta = 1$ so that $\gamma = 2$. In order to look for an inviscid solution, we would set (70) equal to zero and perform the integration, from which we would find that the physical component of vorticity must remain constant along any streamline.

## 4. Invariants

### 4.1. Kinetic Energy

The kinetic energy of a stationary vortex ring with large elliptical cross section is:

$$E = \frac{1}{2}\rho \int_{\bar{V}} \bar{v}_2 \bar{v}^2 \sqrt{g}\, d\bar{V} = \frac{1}{2}\rho \int_0^{\bar{x}^1_{\max}} \int_0^{2\pi} \int_0^{2\pi} \sqrt{g_{22}} \bar{v}^2 \bar{v}^2 \sqrt{g}\, d\bar{x}^3 d\bar{x}^2 d\bar{x}^1$$

$$= \pi\rho \int_0^{\bar{x}^1_{\max}} \int_0^{2\pi} \sqrt{g_{22}} \bar{v}^2 \bar{v}^2 \sqrt{g}\, d\bar{x}^2 d\bar{x}^1$$

### 4.2. Impulse

The impulse is given by:

$$I = \frac{1}{2}\rho \int_{\bar{V}} \mathbf{R} \times \boldsymbol{\omega}\, \sqrt{g}\, d\bar{V} = \frac{1}{2}\rho \int_0^{\bar{x}^1_{\max}} \int_0^{2\pi} \int_0^{2\pi} R\, \bar{\omega}(3) \sqrt{g}\, d\bar{x}^3 d\bar{x}^2 d\bar{x}^1$$

$$= \pi\rho \int_0^{\bar{x}^1_{\max}} \int_0^{2\pi} R\sqrt{g_{33}} \bar{\omega}_A^3 \sqrt{g}\, d\bar{x}^2 d\bar{x}^1$$

$$= \pi\rho \int_0^{\bar{x}^1_{\max}} \int_0^{2\pi} RR\, \bar{\omega}_A^3 \sqrt{\bar{g}}\, d\bar{x}^2 d\bar{x}^1$$

### 4.3. Circulation

The circulation can be found from Eq. (25):

$$\Gamma = \bar{\omega}_A^3\, \bar{S}_3(\bar{x}^1_{\max})$$

where $\bar{S}_3(\bar{x}^1_{\max}) = \pi A B^2 (\bar{x}^1_{\max})^3 c_{\max} = \pi ab(R_O + R_I)/2$. By combining this value with Eq. (35), we have:

$$\Gamma = \frac{8 v_O R_O (R_O - R_C)[P(\bar{x}^1_{\max}) + Q(\bar{x}^1_{\max})]}{a^2 B^2 (R_O^2 - R_I^2)}\, \pi\, ab\, \frac{(R_O + R_I)}{2}$$

$$\therefore \Gamma = \frac{4\pi\, \bar{x}^1_{\max} v_O R_O (R_O - R_C)[P(\bar{x}^1_{\max}) + Q(\bar{x}^1_{\max})]}{aB(R_O - R_I)} \tag{72}$$

### 4.4. Examples

The method of non-dimensionalizing the invariants depends upon the way in which the vortex ring is generated. For a vortex ring generated by an axisymmetric jet in a quiescent fluid, Mohseni and Gharib (1998) show two methods for non-dimensionalizing the kinetic energy and the circulation. The first method involves Norbury's mean core radius, $\alpha$:

$$E_N = \frac{E}{(\omega\alpha)^2 R_C^5} \qquad \Gamma_N = \frac{\Gamma}{\omega\alpha R_C^2},$$

where $\omega$ is the constant vorticity tensor in spherical coordinates. The second method uses combinations of the invariants, as follows:

$$E_{nd} = \frac{E}{\sqrt{\rho I \Gamma^3}} \qquad \Gamma_{nd} = \frac{\Gamma}{I^{1/3} U_P^{2/3}}.$$

The dimensionless kinetic energy above is essentially the same as that used by Gharib *et al.* (1998) and by Mohseni *et al.* (2001).

For vortex rings that form in axisymmetric wake flows, the following formulae have been used to non-dimensionalize the invariants (see, e.g., Johari and Stein, 2002):

$$\Gamma_W = \frac{\Gamma}{v_\infty D} \qquad E_W = \frac{E}{\rho v_\infty^2 D^3} \qquad I_W = \frac{I}{\rho v_\infty D^3},$$

where $v_\infty$ is the free stream velocity and $D$ is the diameter of the axisymmetric body. The subscript "$W$" signifies that the vortex ring is formed by a wake.

Figure 6 shows various geometrical parameters of a family of vortex rings with common values of $R_C$, $k$, $I$, and $E/\Gamma$ but with varying axis ratio. The integrations for the kinetic energy and the impulse were performed numerically using the Composite Simpson's method. Figure 7 shows a family of vortex rings with the axis ratio held fixed at $a/b = 1.5$ and the vortex centre, $R_C$, allowed to vary. Figure 8 shows the vortex geometry and corresponding velocity magnitude at $z = 0$ for a vortex ring roughly comparable to one studied by Gharib *et al.* (1998).

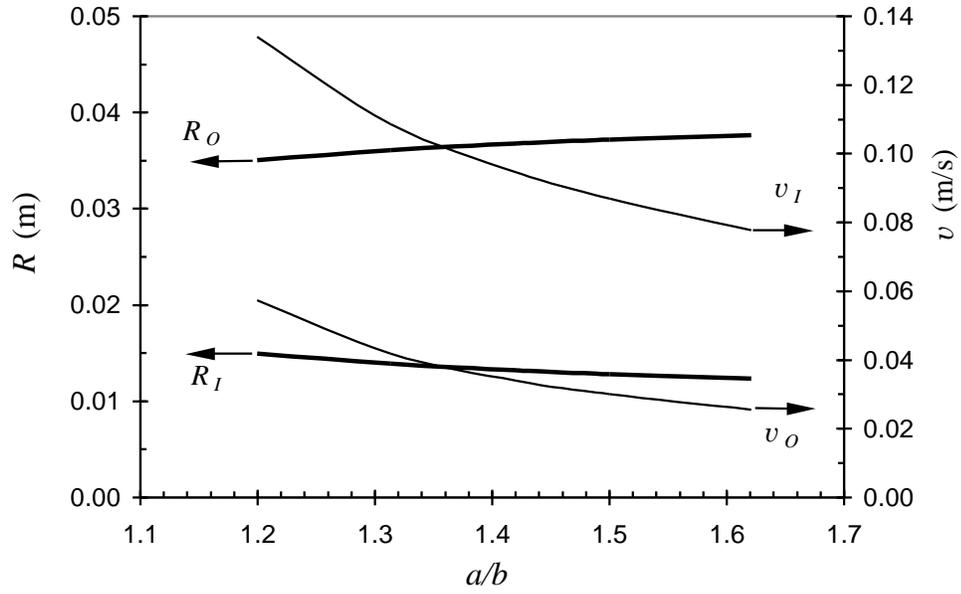

Figure 6. Geometrical parameters of a family of vortex rings with $R_C = 0.025\,\text{m}$, $k = 0$, $I = 0.0376\,\text{N}\cdot\text{s}$, $E/\Gamma = 0.104$, $\rho = 1000\,\text{kg/m}^3$.

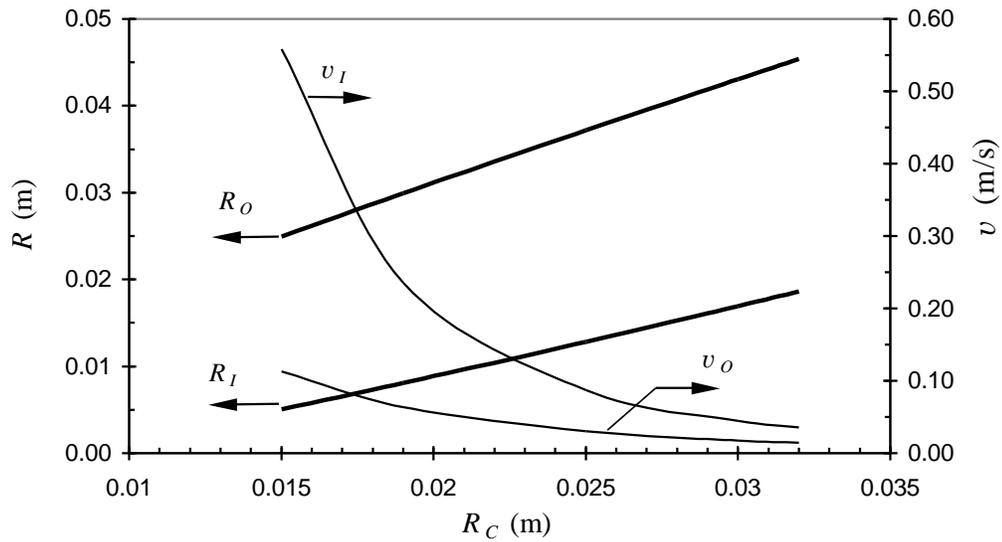

Figure 7. Family of vortex rings with $I = 0.0376\,\text{N}\cdot\text{s}$, $E/\Gamma = 0.104$, $\rho = 1000\,\text{kg/m}^3$, $a/b = 1.5$, and $k = 0$.

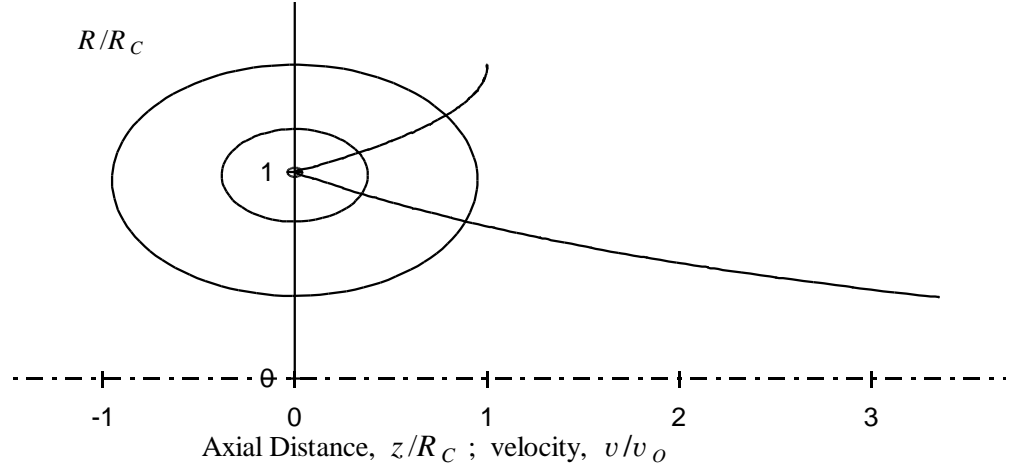

Figure 8. Vortex ring and velocity profile with $v_O = 0.04\,\text{m/s}$, $R_C = 0.027\,\text{m}$, $a/b = 1.7$, $\alpha = 0.7295$, $\Gamma = 80\text{ cm}^2/\text{s}$, $E = 0.0082\,\text{J}$, $I = 0.0212\,\text{N}\cdot\text{s}$, $\Gamma_{nd} \approx 0.112$, $E_{nd} = 2.47$, $\Gamma_N = 0.0594$, $E_N = 0.0166$, $I_N = 0.215$, $\Gamma_W \approx 3.7$, $E_W \approx 33$, $I_W \approx 3.4$, $\rho = 1000\text{ kg/m}^3$. Assumptions: piston velocity $U_P \approx v_I$; free stream velocity $v_\infty \approx v_O$.

## 5. Application to the Strouhal Number

The vortex shedding period in unsteady wake flows is the same as the period of the time-mean toroidal vortex resulting from conditional sampling over one shedding cycle. This period can be determined by integrating (36), as follows:

$$\int_0^{2\pi} \sqrt{\bar{g}}\,\mathrm{d}\bar{x}^2 \;=\; v_O R_O B(1-k)\left(\frac{\bar{x}^1}{\bar{x}^1_{\max}}\right)^3 \frac{c}{c_{\max}}\frac{[P(\bar{x}^1_{\max}) + Q(\bar{x}^1_{\max})]}{[P(\bar{x}^1) + Q(\bar{x}^1)]}\int_0^T \mathrm{d}t$$

$$\pi(2R_C - 3\bar{x}^1 Bk)v \;=\; \frac{v_O R_O(1-k)}{\bar{x}^1 A}\left(\frac{\bar{x}^1}{\bar{x}^1_{\max}}\right)^3 \frac{c}{c_{\max}}\frac{[P(\bar{x}^1_{\max}) + Q(\bar{x}^1_{\max})]}{[P(\bar{x}^1) + Q(\bar{x}^1)]}T$$

Since the shedding frequency, $f_S$, is related to the period of the time-mean vortex by: $f_S = m/T$, where $m$ is the number of vortices shed downstream per shedding cycle, we have:

$$f_S \;=\; \frac{m v_O R_O}{2\pi A R_C}\frac{\bar{x}^1}{(\bar{x}^1_{\max})^2}\frac{[P(\bar{x}^1_{\max}) - c(\bar{x}^1_{\max})]}{[P(\bar{x}^1) - c(\bar{x}^1)]}. \tag{73}$$

Here some simplification has been gained by assuming $k = 0$.

A Strouhal number can be defined by:

$$S_O \;=\; \frac{f_S(R_O - R_C)}{v_O} \;=\; \frac{mR_O(R_O - R_C)}{2\pi a R_C}\left(\frac{\bar{x}^1}{\bar{x}^1_{\max}}\right)\frac{[P(\bar{x}^1_{\max}) - c(\bar{x}^1_{\max})]}{[P(\bar{x}^1) - c(\bar{x}^1)]}. \tag{74}$$

Notice that the Strouhal number is independent of the direction, $\bar{x}^2$, and should, therefore, remain constant along streamlines within the core. It varies in the $\bar{x}^1$ direction and can be evaluated at the axial station coincident with the vortex centre by letting $\bar{x}^2 = \pm \pi/2$. Along the line $\bar{x}^2 = \pi/2$, the coordinate $\bar{x}^1$ is given by:

$$\bar{x}^1 = \frac{b}{B}\frac{(R-R_C)}{(R_O-R_C)} \qquad (\bar{x}^2 = \pi/2),$$

and when $\bar{x}^2 = -\pi/2$:

$$\bar{x}^1 = \frac{b}{B}\frac{(R_C-R)}{(R_C-R_I)} \qquad (\bar{x}^2 = -\pi/2).$$

Simplifying, (74) becomes:

$$S_O \equiv \frac{f_S(R_O-R_C)}{v_O} = \frac{mR_O|R-R_C|}{2\pi\, aR_C}\frac{[P(\bar{x}^1_{\max})-c(\bar{x}^1_{\max})]}{[P(\bar{x}^1)-c(\bar{x}^1)]}. \tag{75}$$

Values of this Strouhal number at $z=0$ are shown in Figure 9 for the vortex rings shown in Figures 3, 4, and 5. The Strouhal number defined by (75) is related to the classical definition ($St = f_S D/v_\infty$) as follows:

$$St = S\frac{v_O}{v_\infty}\frac{D}{(R_O-R_C)}. \tag{76}$$

The grouping defined by (75) may be of use in vortex shedding studies of axisymmetric flows.

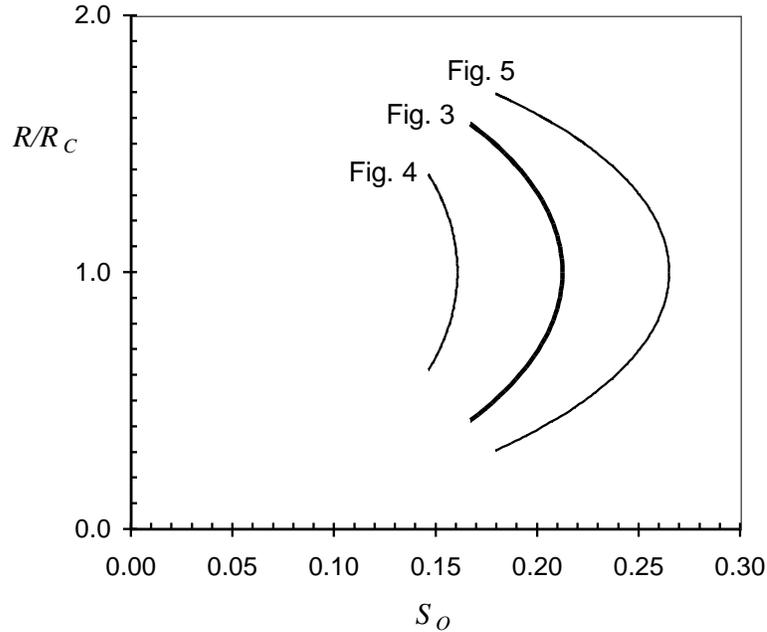

Figure 9. Strouhal number defined by (75) versus radial distance from axis of symmetry at $z=0$ for the vortex rings in Figures 3, 4 and 5. ($m=1$)

A Strouhal number can also be formulated in terms of properties at the inner radius rather than the outer radius. From (65) we know that $v_O R_O = v_I R_I$; therefore, these quantities can be interchanged in (73), giving:

$$S_I \equiv \frac{2f_S R_I}{v_I} = \frac{mR_I{}^2 |R - R_C|}{\pi a R_C (R_O - R_C)} \frac{[P(\bar{x}^1_{\max}) - c(\bar{x}^1_{\max})]}{[P(\bar{x}^1) - c(\bar{x}^1)]} \tag{77}$$

This grouping would correspond roughly to $f_S D / v_J$, where $D$ is the jet diameter and $v_J$ is the jet velocity.

## 6. Conclusion

Hill's vortex was discussed and shown to be a spherical region of uniform vorticity. An explicit algebraic expression was found for the velocity field within a steady, toroidal vortex with large elliptical cross section. The vorticity decreases monotonically with distance from the symmetry axis. As the inner radius of the torus approaches zero, the velocity of the central jet increases without bound due to the infinitely small flow area available for the reverse flow. This does not occur in the spherical vortex because the bounding streamline contains two points of zero velocity, namely the forward and rear stagnation points. The toroidal vortex appears to be more suitable for modelling centrally-driven (jet) flows, whereas the spherical vortex may be more suitable for modelling externally-driven (wake) flows. A general formulation for the stream function is presented that satisfies the continuity equation in coordinate systems for which a certain statement (Eq. (46)) can be made about the flow in one of the coordinate directions. A Strouhal number formulation is presented based on the solution for the velocity field within the time-mean vortex ring.